\documentclass[conf]{new-aiaa}
\usepackage[utf8]{inputenc}

\usepackage{graphicx}
\usepackage{amsmath}
\usepackage[version=4]{mhchem}
\usepackage{siunitx}
\usepackage{longtable,tabularx}
\usepackage{subcaption}
\usepackage{rotating}
\usepackage{multirow}
\usepackage{hyperref}
\usepackage{graphicx} 
\usepackage{pgfplots}
\usepackage[capitalise]{cleveref}
\title{A Reinforcement Learning Approach to Quiet and Safe UAM Traffic Management}

\author{Surya Murthy\footnote{Graduate Research Assistant, Oden Institute for Computational Engineering and Sciences, AIAA Student Member}, John-Paul Clarke\footnote{Professor and Ernest Cockrell, Jr. Memorial Chair in Engineering, Department of Aerospace Engineering and Engineering Mechanics, AIAA Fellow}, Ufuk Topcu\footnote{Professor and W.A. “Tex” Moncrief, Jr. Chair in Computational Engineering and Sciences VI, Oden Institute for Computational Engineering and Sciences, AIAA Senior Member}}
\affil{The University of Texas at Austin, Austin, Texas 78712}

\author{Zhenyu Gao\footnote{Assistant Professor, Department of Mechanical and Aerospace Engineering, AIAA Member}}
\affil{The Hong Kong University of Science and Technology, Clear Water Bay, Hong Kong}

\begin{document}

\maketitle

\begin{abstract}

Urban air mobility (UAM) is a transformative system that operates various small aerial vehicles in urban environments to reshape urban transportation. However, integrating UAM into existing urban environments presents a variety of complex challenges. Recent analyses of UAM's operational constraints highlight aircraft noise and system safety as key hurdles to UAM system implementation. Future UAM air traffic management schemes must ensure that the system is both quiet and safe. We propose a multi-agent reinforcement learning approach to manage UAM traffic, aiming at both vertical separation assurance and noise mitigation. Through extensive training, the reinforcement learning agent learns to balance the two primary objectives by employing altitude adjustments in a multi-layer UAM network. The results reveal the tradeoffs among noise impact, traffic congestion, and separation. Overall, our findings demonstrate the potential of reinforcement learning in mitigating UAM's noise impact while maintaining safe separation using altitude adjustments. 

\end{abstract}

\section{Introduction}

Urban air mobility (UAM) is a transformative system that has the potential to revolutionize urban transportation and support various urban services, including passenger mobility, cargo delivery, infrastructure monitoring, public safety, and emergency response. UAM is anticipated to improve public welfare, especially in underserved local and regional communities. While UAM offers substantial opportunities for enhancing urban services and promoting public welfare, challenges such as community noise acceptance, safety concerns, and infrastructure requirements hinder its successful integration into existing urban settings ~\cite{barsotti2024eco}. Since UAM operates much closer to urban residents than traditional air transport systems, addressing the noise that UAM aircraft produce is crucial, as it can pose health risks like sleep disturbances and a higher incidence of cardiovascular issues for residents near flight operations. For mitigating noise in UAM, adopting noise-aware flight operations planning is vital, regardless of the aircraft noise reduction technologies available~\cite{gao2024noise}.

Maintaining separation between aircraft is critical to preventing collisions and ensuring system safety in UAM operations. Traditional air traffic management approaches often rely on deterministic algorithms for conflict detection and resolution across various time horizons. These methods are effective but may not account for modern challenges, such as sustainability concerns. Recent advancements in reinforcement learning (RL) have introduced models capable of enforcing safe separation by dynamically adjusting aircraft speeds and learning complex interactions between vehicles ~\cite{DENIZ2024100157, brittain2020deep, zhao2021physics}. However, these models typically focus solely on safety, neglecting sustainability aspects like noise impact on communities. Our work addresses this limitation by incorporating noise mitigation as a parallel objective, enabling dynamic altitude adjustments that balance safe separation with reduced noise impact.


Given the importance of safe and quiet UAM operations, we propose RL techniques to actively reduce noise impact in real time at the vehicle level. We formulate noise mitigation and vertical separation assurance as a multi-agent reinforcement learning problem and define the UAM environment as a Markov decision process (MDP). Within this framework, we model noise impact and vertical separation assurance as objectives within the reward function. Through training, the reinforcement learning agent learns to balance these objectives effectively, employing altitude adjustments to reduce noise impact while using altitude adjustments to maintain a safe vertical separation from other aircraft. By enforcing vertical separation between aircraft, we reduce the likelihood of general loss of separation (LOS) events, which also consider horizontal proximity. We find that a trained RL policy can use altitude adjustments to reduce cumulative noise impact while maintaining safe separation between other aircraft. Our results highlight a tradeoff between noise reduction and vertical separation in UAM operations. As aircraft prioritize higher altitudes to minimize ground noise, we observe an increase in air traffic congestion and the frequency of LOS events. Conversely, encouraging a wider distribution of altitudes reduces LOS events but raises noise impact due to lower-altitude operations. This tradeoff emphasizes the need for carefully balanced policies for UAM operations to achieve both environmental sustainability and operational safety.

\section{Background and Literature Review}\label{sec:lit}
\subsection{Safe Operations in Aviation}
Conflict detection methods play a central role in ensuring air traffic safety by identifying and resolving potential conflicts over short time horizons. Systems like ACAS X (Airborne Collision Avoidance System X) use probabilistic models to forecast aircraft trajectories and dynamically generate avoidance maneuvers \cite{manfredi2016introduction}. Similarly, reachability analysis \cite{tomlin1998conflict} and trajectory optimization \cite{farley2007automated} establish future state predictions and compute safe paths to prevent conflicts. These methods operate over shorter timeframes and typically resolve conflicts within 2–3 minutes. In contrast, our approach aims to address safety concerns over longer time horizons, focusing on congestion reduction at various altitude levels. By proactively managing traffic flow, we aim to alleviate the workload on short-time-horizon conflict detection systems, allowing them to function more efficiently within reduced-congestion environments.

Strategic separation management uses strategic decisions like ground delays made by air traffic managers to balance traffic demand with airspace capacity at bottlenecks. 
These bottlenecks include airport runways, merging points, and air route intersections.
Strategic separation management plays an important role in modern air traffic management, with programs such as the Ground Delay Program ~\cite{odoni1987flow} and Airspace Flow Program ~\cite{libby2005operational} serving as key examples.
Demand-capacity balancing (DCB) approaches use optimization and heuristic techniques to schedule flight takeoffs to balance demand with available capacity at bottlenecks~\cite{chen2024integrated}.
By integrating with real-time deconfliction, strategic separation approaches effectively enforce safety constraints while maintaining operational efficiency.
While strategic separation primarily manages long-term demand, our approach aims to operate in real-time, allowing it to complement strategic separation by dynamically adapting to unforeseen fluctuations in traffic.

Tactical separation approaches address real-time conflict resolution by enforcing safe separation between aircraft, often utilizing techniques like Monte-Carlo Tree Search \cite{wu2022safety} and A* path planning \cite{zhao2021multiple}. Recently, researchers have made significant progress in applying deep reinforcement learning (DRL) methods for tactical separation assurance. DRL approaches model the UAM environment as a MDP and train aircraft to avoid conflicts using only local observations. Various DRL models incorporate advanced architectures, including long-short term memory networks \cite{DENIZ2024100157}, attention layers \cite{brittain2020deep}, and physics-based models \cite{zhao2021physics}. Most of these models resolve conflicts and reduce LOS events by enforcing planar separation between aircraft. In contrast, our work focuses on enforcing vertical separation, which reduces the likelihood of general LOS events by ensuring safe vertical distances between aircraft. Building on this foundation, we expand the focus by integrating noise mitigation as an additional objective, a critical consideration in UAM where aircraft operate close to populated areas. Our approach enables the DRL agent to dynamically adjust trajectories to minimize noise exposure while maintaining safety, introducing a multi-objective framework for tactical vertical separation that balances both environmental and safety considerations in densely populated settings.

\subsection{Urban Air Mobility Noise Modeling and Mitigation}

While UAM has significant potential to enhance urban services and public welfare, its integration into existing urban environments presents several complex challenges. Lessons learned from past urban helicopter services, along with recent evaluations of UAM's operational constraints, emphasize that community acceptance of aircraft noise, system safety, and infrastructure development are critical hurdles to the successful implementation and expansion of UAM systems~\cite{vascik2017constraint,vascik2018analysis,pons2022understanding}. Limiting the noise impact of UAM aircraft on urban communities is especially crucial for a sustainable UAM system. Since UAM operates in close proximity to urban populations, the noise generated by UAM aircraft poses significant health risks, including psychological distress, sleep disturbances, and an increased likelihood of cardiovascular diseases~\cite{gao2022multi}. Overall, the reduction of UAM noise has emerged as a key priority within the aviation and aerospace communities.

Noise mitigation strategies for UAM are built on two key research pillars: aircraft noise modeling and noise-aware operational planning. At present, noise modeling predominantly uses acoustic simulations, largely due to the scarcity of measurement data for new eVTOL aircraft configurations~\cite{rizzi2022prediction,rizzi2023modeling,bian2021assessment}. To effectively mitigate noise in aviation, noise-aware flight operations planning is an essential approach, regardless of the aircraft noise reduction technologies available. In commercial aviation, noise abatement strategies such as the continuous descent approach (CDA)~\cite{clarke2004continuous} and noise abatement departure procedures (NADP)~\cite{lim2020noise} have been the subject of extensive research. In the UAM sector, most research efforts have focused on noise-aware flight trajectory planning and airspace management for both eVTOL aircraft and small drones. The objective of noise-aware flight trajectory planning is to generate 2-D or 3-D flight trajectories that steer clear of people and buildings on the ground~\cite{gao2023noise}, while also considering population information~\cite{woodcock2022development}. The complexity inherent in the noise patterns of UAM aircraft poses challenges for trajectory optimization, leading the majority of research to rely on heuristic algorithms~\cite{pang2022uav,tan2024enhancing}. In contrast, advanced techniques such as convex optimization and nonconvex optimal control are less frequently explored~\cite{wu2024convex}. Airspace management~\cite{bauranov2021designing} represents an alternative strategy for lessening the community noise impact associated with UAM systems. The core concept involves transforming various noise constraints present in urban areas into designated no-fly zones for UAM operations, thereby minimizing the overall noise disturbance to the community~\cite{gao2023noise}.

A challenge for most existing UAM noise mitigation solutions is their compatibility and integration into current ATM schemes. Recognizing the effectiveness of RL approaches in improving the safety and efficiency of UAM operations, it is worthwhile to investigate an RL-based solution for mitigating the noise impact of UAM. Our proposed approach represents one of the first attempts to leverage RL techniques for managing UAM noise while simultaneously considering vertical separation assurance within the same framework.

\section{Methodology}\label{sec:method}

This section describes the components of the proposed methodology, including (i) a brief overview of the UAM flight network and operations, (ii) an aircraft noise model, (iii) our formulation of the UAM network as a MDP with vertical separation assurance and noise management as objectives in the reward function, and (iv) the reinforcement learning model we use to balance these objectives.

\subsection{Urban Air Mobility Network and Operations}

The UAM system is projected to operate within a network consisting of vertiports, designated areas for aircraft takeoff and landing, and flight corridors, which are the routes aircraft follow when traveling between vertiports \cite{faaconops2023}. Aircraft travel between an origin and a destination via a series of flight corridors and cannot deviate from their route by changing flight corridors. The primary objectives of each aircraft are to minimize noise impact and maintain safe separation from other aircraft. Loss of separation (LOS) events occur whenever two aircraft come within \(d_{\text{LOS}}\) of each other, measured as the Euclidean distance between them, which accounts for both horizontal and vertical separation.

\begin{figure}[h!]
     \centering
     \begin{subfigure}[b]{0.45\textwidth}
         \centering
         \includegraphics[width=\textwidth]{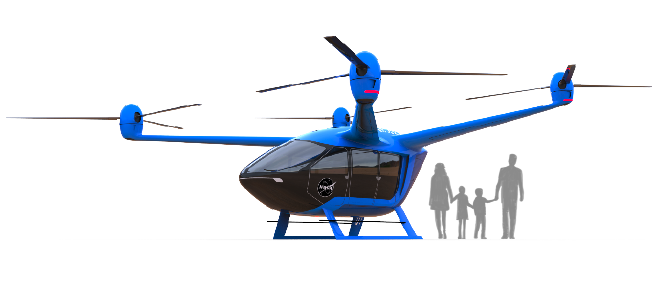}
         \caption{The NASA RVLT quadrotor vehicle concept}
         \label{fig:nasavehicle}
     \end{subfigure}
     \hspace{0.5cm}
     \begin{subfigure}[b]{0.475\textwidth}
         \centering
         \includegraphics[width=\textwidth]{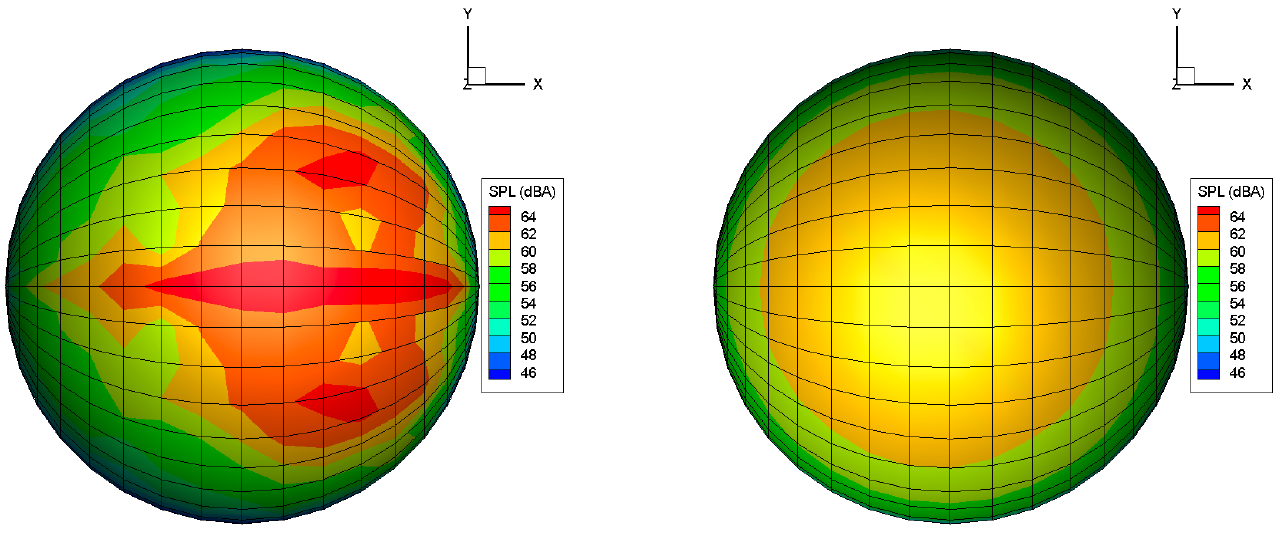}
         \caption{The example noise hemispheres from acoustic simulation}
         \label{fig:noisesphere}
     \end{subfigure}
        \caption{The configuration of the NASA RVLT vehicle and an example of its noise simulation (sources: \cite{rizzi2022prediction,rizzi2023modeling})}
        \label{fig:vehicle}
\end{figure}

\subsection{Aircraft Noise Model}

An aircraft noise model is a crucial component for planning noise-aware UAM operations. The specific approach to modeling aircraft noise depends on various factors, including modeling range and fidelity, computational time, and urban terrain. The assessment of UAM aircraft's noise impact on urban communities often relies on a standard known as noise-power-distance (NPD) data~\cite{gao2024noise}. NPD data describes the relationship between the noise impact of an aircraft and its slant distance to the receiver under various conditions. Currently, NPD data is available for around 300 fixed-wing aircraft and 26 helicopter types that are operated around the world~\cite{volpe2022uam}. The NPD data for each aircraft type is derived from either noise certification tests or controlled tests performed in accordance with stringent international standards. Due to the absence of NPD data for new UAM aircraft configurations like eVTOL aircraft, practitioners have largely depended on acoustic simulations to model the noise patterns (e.g., 3D directivity) associated with these UAM configurations~\cite{rizzi2022prediction}. The results of the simulation are subsequently transformed into NPD data for the effective modeling of UAM's community noise exposure.

We utilize the noise models for the NASA Revolutionary Vertical Lift Technology (RVLT) quadrotor reference vehicle, as illustrated in Figure~\ref{fig:nasavehicle}. Researchers at NASA Langley Research Center~\cite{rizzi2022prediction,rizzi2023modeling} have employed ANOPP2's Aeroacoustic ROtor Noise (AARON) tool to generate noise hemispheres for this reference aircraft configuration, depicted in Figure~\ref{fig:noisesphere}. This noise model is used in conjunction with flight simulations to produce Sound Exposure Level (SEL) NPD data for three typical operational modes: level flyover (L), (dynamic) departure (D), and approach (A).

\begin{figure}[htbp]
	\centering
        \includegraphics[width=0.425\textwidth]{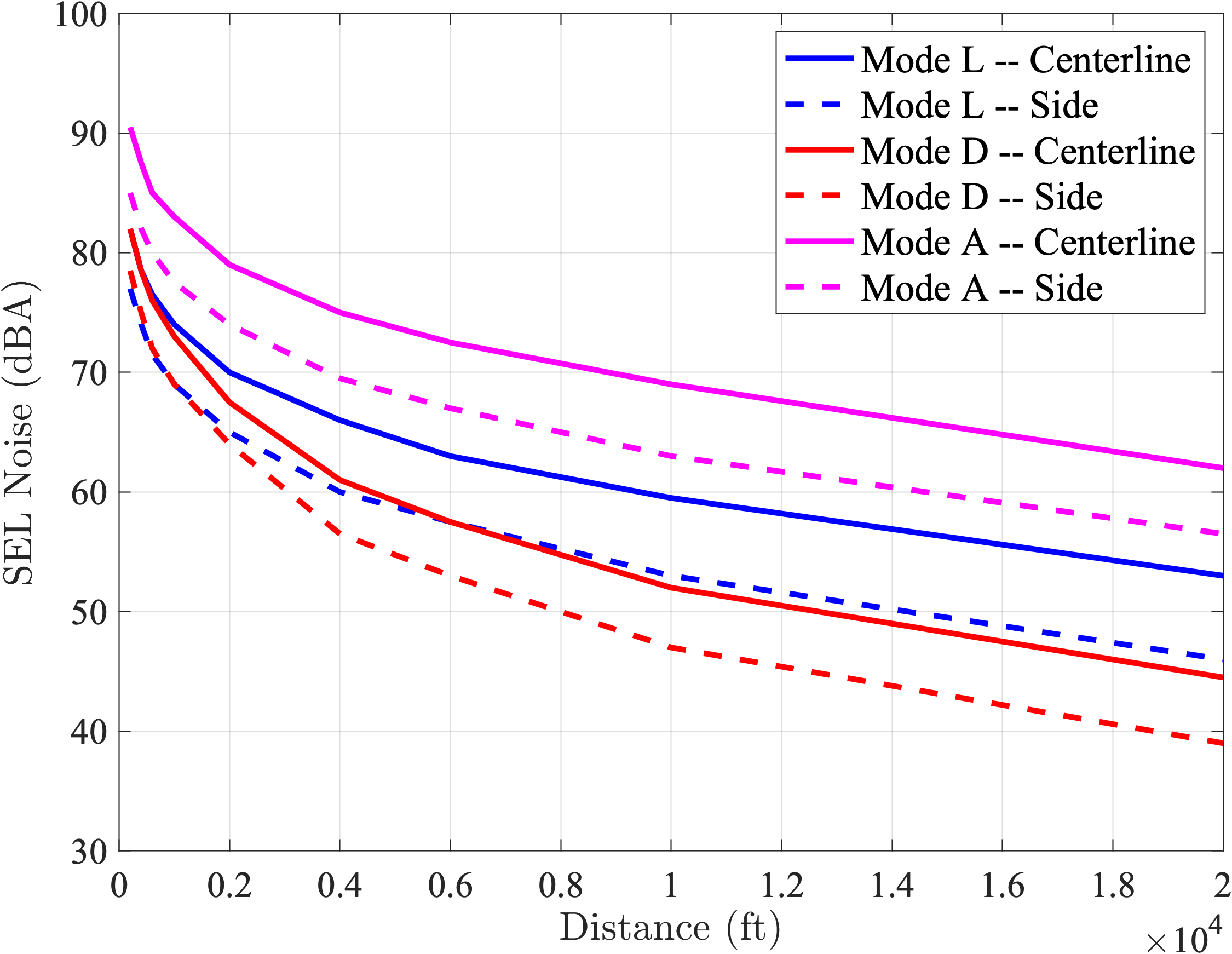}
        \hspace{1cm}
        \includegraphics[width=0.425\textwidth]{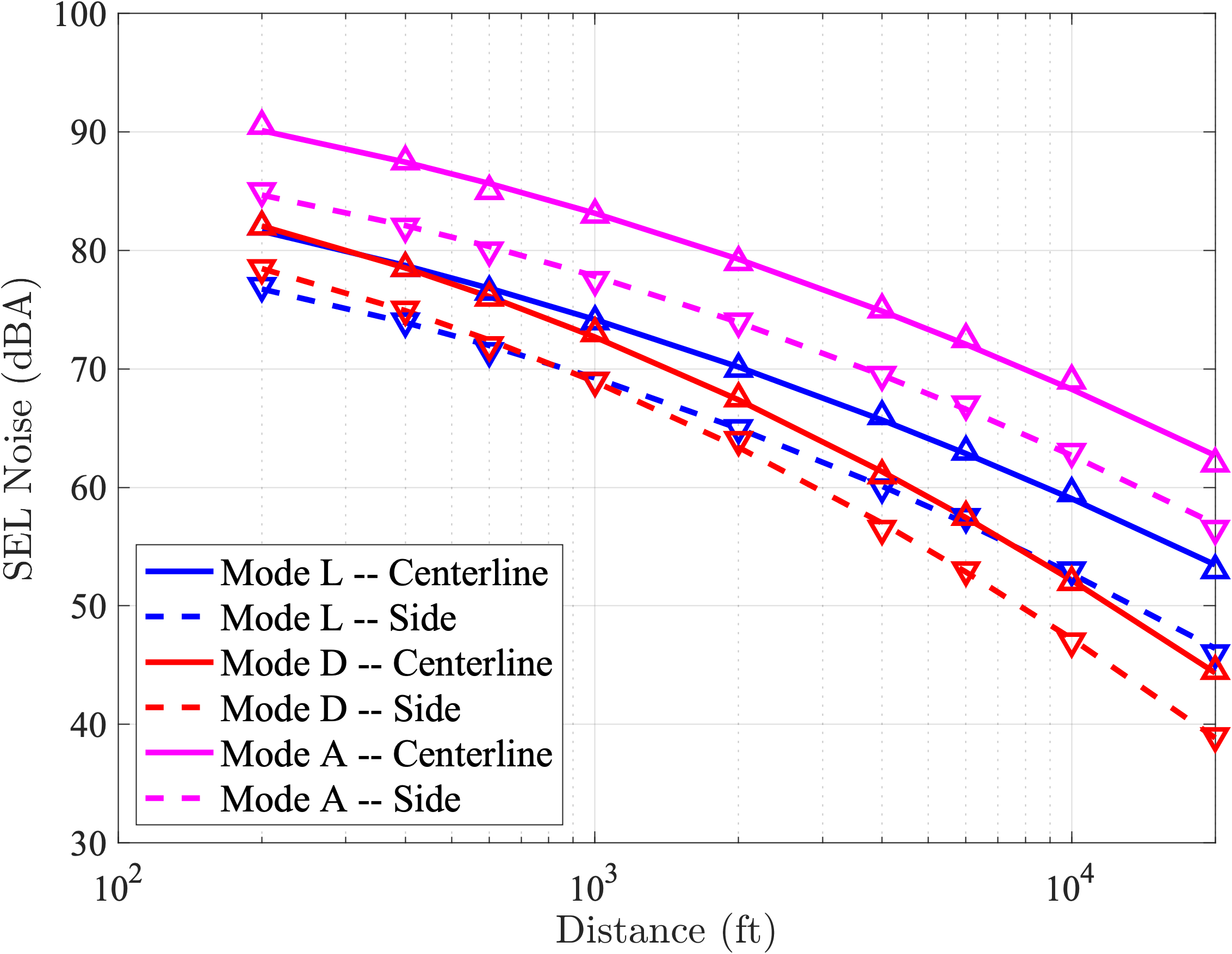}
	\caption{NPD data for the NASA RVLT quadrotor vehicle: normal distance scale (left) and log distance scale with the fitted model (right)}
	\label{fig:npds}
\end{figure}

Figure~\ref{fig:npds} presents the resulting NPD data for the NASA RVLT quadrotor vehicle. Due to the availability of NPD data only at specific distance levels ranging from 200 to 20,000 ft, we employ logarithmic interpolation to estimate noise levels at distances not directly measured. We fit regression models to the NPD data illustrated in the left plot of Figure~\ref{fig:npds}. For each condition, which consists of a specific combination of operational mode and measurement position, the regression model for single-event noise has the following form:

\begin{equation}\label{eq:Rn}
    N_{\text{Single}} = c_0 + c_1 \log_{10}{z} + c_2 (\log_{10}{z})^2
\end{equation}
where $N_{\text{Single}}$ is noise in A-weighted SEL and $z \in [200, 20000]$ is distance in ft. The right plot of Figure~\ref{fig:npds} displays the goodness-of-fit for the six NPD curves, with their corresponding regression coefficients presented in Table~\ref{tbl:fittednpds}.
\begin{table}[htbp]
    \centering
    \caption{Regression coefficients for the NPD curves}
    \begin{tabular}{llll}
    \hline
    Condition       & $c_0$    & $c_1$    & $c_2$    \\ \hline
    Mode L - Centerline & 88.09 & 3.21  & -2.62 \\ 
    Mode L - Side   & 78.01 & 7.26  & -3.39 \\ 
    Mode D - Centerline & 84.05 & 8.76  & -4.18 \\ 
    Mode D - Side   & 77.34 & 11.34 & -4.72 \\ 
    Mode A - Centerline & 93.35 & 5.17  & -2.86 \\ 
    Mode A - Side   & 85.55 & 6.83  & -3.14 \\ \hline
    \end{tabular}
    \label{tbl:fittednpds}
\end{table}

In this study, to simulate the noise impact of UAM operations during flyover, we primarily utilize the regression coefficients from the condition `Mode L - Centerline', found in the first row of Table~\ref{tbl:fittednpds}. Finally, the cumulative noise increase over the ambient noise level $N_{\text{ambient}}$ of $N$ aircraft over a given airspace is defined as
\begin{equation}
    \begin{aligned}
        N_{\text{increase}} &= N_{\text{cumulative}} - N_{\text{ambient}}\\
        &= 10\log_{10}( \sum_{i = 1}^{N}10^{N_{\text{Single}}^i /10}) -  35.56  - N_{\text{ambient}}
    \label{eq:cumulative}
    \end{aligned}
\end{equation}

\subsection{Urban Air Mobility Network as a Markov Decision Process}

Markov Decision Processes (MDPs) provide a mathematical framework for modeling decision-making problems where an agent interacts with an environment in discrete time steps \cite{sutton2018reinforcement}. An MDP is defined by the following elements:
\begin{itemize}
    \item \textbf{State space ($S$)}: The set of all possible states the agent can encounter. A state represents the configuration or condition of the system at a given time.
    \item \textbf{Action space ($A$)}: The set of all possible actions the agent can take. An action represents a choice made by the agent that influences the environment.
    \item \textbf{Transition function ($T(s, a)$)}: The probability distribution over the next state $s'$ given the current state $s$ and action $a$. This describes how the environment changes in response to the agent's actions.
    \item \textbf{Reward function ($R(s, a)$)}: The immediate reward the agent receives after taking action $a$ in state $s$. The reward function provides feedback to the agent about the desirability of its actions.
\end{itemize}
An MDP provides a mathematical framework for modeling sequential decision-making. Agents interacting within an MDP environment exist in a state ($s \in S$), take an action ($a \in A$), transition to a new state ($s' = T(s, a)$), and receive a reward ($r = R(s, a)$). We model the Urban Air Mobility (UAM) network as an MDP, where the state space, action space, and reward functions are defined to address noise and safety constraints in urban environments. We use the BlueSky Air Traffic Simulator to simulate the UAM environment \cite{Simulator}. The transition function is handled by the simulator's built-in dynamics, which are not explicitly defined here.

\subsubsection{State and Action Space}

The state contains the information necessary for an agent to make decisions. To address the noise impact of UAM on urban communities, our state space includes factors related to noise impact and safety considerations, such as vertical separation. The state space of each aircraft at time step \(t\) is defined as
\begin{equation}
    s_{t} = \{z, b_{\text{changing}}, z_{\text{target}}, a_{t-1}\},
\end{equation}
where \(z\) is the aircraft's current altitude (in feet), \(b_{\text{changing}} \in \{0, 1\}\) is a boolean indicating whether the aircraft is in the process of changing altitude, \(z_{\text{target}}\) is the target altitude, and \(a_{t-1}\) is the previous action taken by the aircraft.

To enforce safe vertical separation constraints between aircraft, it is also necessary to include the state information of neighboring aircraft in the state space. We assume that each aircraft receives broadcasted state information from other aircraft within a specified communication range. The state of a neighboring aircraft \(i\) is defined as:

\begin{equation}
    h_t(i) = \{z_{\text{rel}}^{(i)}, d_o^{(i)}, a_{t-1}^{(i)}\},
\end{equation}
where \(z_{\text{rel}}^{(i)}\) is the relative altitude of the neighboring aircraft, \(d_o^{(i)}\) is the distance between the aircraft and the neighboring aircraft, and \(a_{t-1}^{(i)}\) represents the neighboring aircraft’s previous action. This shared information enables each aircraft to ensure safe vertical separation from other aircraft, thereby contributing to collision avoidance.

The action space for aircraft consists of altitude change advisories, which we partition into discrete altitude levels to maintain safe vertical separation. Formally, the action space is defined as:

\begin{equation}
    A = \{v_{z0}, v_{z-}, v_{z+}\},
\end{equation}
where \(v_{z+}\) and \(v_{z-}\) represent actions to increase and decrease the aircraft's altitude, respectively, while \(v_{z0}\) corresponds to maintaining the current altitude. A critical design constraint is that once an aircraft decides to change altitude, it must complete its altitude adjustment to the target level \(z_{\text{target}}\) before it can execute another altitude change. This rule prevents the model from continuously oscillating between altitude levels, ensuring that the RL agent observes the full consequences of its actions. The inclusion of this constraint is especially important given the relatively slow vertical speed of aircraft, where frequent altitude changes could disrupt learning by preventing the model from capturing the effects of each action over time.

\subsubsection{Reward Function}

The SEL model defined in \cref{eq:Rn} is incorporated into the noise reward function. The noise exposure reward function, \(R_{\text{noise}}\), is defined as
\begin{equation} \label{eq:Rs}
    R_{\text{noise}}(s_{t}) = \frac{N_{\text{Single}} - N_{\text{min}}}{N_{\text{max}} - N_{\text{min}}},
\end{equation}
where \(N_{\text{max}} = c_0 + c_1 \log_{10}{(z_{\text{min}})} + c_2 (\log_{10}{(z_{\text{min}})})^2\) and \(N_{\text{min}} = c_0 + c_1 \log_{10}{(z_{\text{max}})} + c_2 (\log_{10}{(z_{\text{max}})})^2\) represent the aircraft’s maximum and minimum individual noise impact, respectively. This reward function represents the normalized noise impact caused by an individual aircraft. By training a RL model using this reward function, each aircraft is incentivized to minimize its noise impact by dynamically adjusting its altitude. Each aircraft minimizing its individual noise impact allows for a reduction of the cumulative noise impact.

Another objective we consider is vertical separation assurance using altitude adjustments. When enforcing a safe separation constraint, we recognize that having multiple aircraft at the same altitude level increases the risk of loss of separation. Therefore, we define the vertical separation cost as

\begin{equation}
R_{\text{separation}}(s_t, h_t) = - \min\left(\lambda \sum_{i=1}^{N} I(z - z^{(i)} < d_{\text{LOS}}), 1\right),
\end{equation}
where \(d_{\text{LOS}}\) represents the minimum safe separation distance between aircraft, and \(N\) is the total number of neighboring aircraft. This reward function penalizes aircraft for flying at the same altitude as neighboring aircraft, with the penalty increasing as the number of nearby aircraft grows. The hyperparameter \(\lambda\) controls the rate at which the penalty increases, and in our implementation, we set \(\lambda = 0.1\). By applying this reward function, aircraft are encouraged to maintain safe vertical separation by flying at different altitudes, thereby reducing the risk of conflict.

The total reward function is expressed as

\begin{equation} \label{eq:Rf}
    R(s_t, h_t) = \rho R_{\text{noise}}(s_t) + (1-\rho) R_{\text{separation}}(s_t, h_t),
\end{equation}
where \(\rho\) controls the tradeoff between vertical separation assurance and noise mitigation. By adjusting \(\rho\), we can control how much the RL model prioritizes each objective.

\subsection{Reinforcement Learning Model}



The goal of reinforcement learning (RL) is to learn a policy $\pi: S \xrightarrow{} \mathcal{P}(A)$, which is a mapping from states to action probabilities. The policy defines the agent's behavior by prescribing the action that the agent should take in each state. The agent's objective is to maximize the total cumulative reward it receives over time, which is formally expressed as the expected return:
\begin{equation}
J(\pi) = \mathbb{E}_{\pi} \left[\sum_{t=0}^{T} \gamma^t R(s_t, a_t)\right],
\end{equation}
where \(t\) denotes the time step, \(s_t\) and \(a_t\) are the state and action at time \(t\), and \(T\) is the total number of time steps. In this equation, the summation represents the cumulative reward over all time steps, weighted by the discount factor \(\gamma \in [0, 1]\), which determines the importance of future rewards. The expectation is taken over the distribution of states and actions generated by following the policy \(\pi\).

An optimal policy $\pi^*$ is a policy that maximizes the expected return starting from any initial state. Formally, an optimal policy satisfies:
\begin{equation}
\pi^* = \arg \max_{\pi} \mathbb{E}_{\pi} \left[\sum_{t=0}^{T} \gamma^t R(s_t, a_t)\right].
\end{equation}
This means that the optimal policy yields the highest cumulative reward over time, considering both the immediate and long-term consequences of the agent’s actions. Learning the optimal policy is the primary objective of an RL agent.

A value function $V(s)$ estimates the expected return from a given state $s$, assuming the agent follows a specific policy $\pi$. It quantifies how favorable a state is in terms of the expected cumulative reward that can be achieved from that state onward:
\begin{equation}
V^{\pi}(s) = \mathbb{E}_{\pi} \left[\sum_{t=0}^{T} \gamma^t R(s_t, a_t) \mid s_0 = s \right].
\end{equation}
The value function gives an agent a measure of the long-term benefit of being in a particular state when following the policy $\pi$. The value function allows the agent to evaluate the desirability of different states under a given policy and prioritize actions that lead to states with higher values.

We choose the D2MAV-A RL model~\cite{brittain2020deep} to manage vertical separation while minimizing noise impact. The model takes as input the state of the aircraft, $s_t$, and the states of all neighboring intruding aircraft, $h_t(i)$, which include key information such as relative positions and altitudes. D2MAV-A utilizes a neural network that incorporates an attention layer to map the set of intruder states to a fixed-size representation, enabling the model to focus on the most relevant intruders at each time step. The model generates two key outputs: the value function $V(s)$, which estimates the expected cumulative reward, and the policy $\pi(s, a)$, which provides a probability distribution over possible altitude adjustment actions. These outputs guide the aircraft in selecting actions that maintain safe separation while minimizing environmental noise. 

The training of the D2MAV-A model follows a centralized learning with a decentralized execution framework. During training, each aircraft collects experience tuples consisting of states, actions, rewards, and next states from multiple simulation steps. These experiences are aggregated and used to update a shared neural network. The model is updated using proximal policy optimization, which constrains policy updates to maintain stability while optimizing performance \cite{schulman2017proximal}. Once trained, the shared policy is distributed to each aircraft, allowing them to operate independently based on local observations. By following this iterative process, D2MAV-A can converge to a policy that balances vertical separation and noise impact. Performance is evaluated through metrics such as LOS events and cumulative noise impact.

\section{Experiments}

\subsection{The South Austin Case Study}

\begin{figure}[hbt!]
     \centering
     \includegraphics[width=0.55\textwidth]{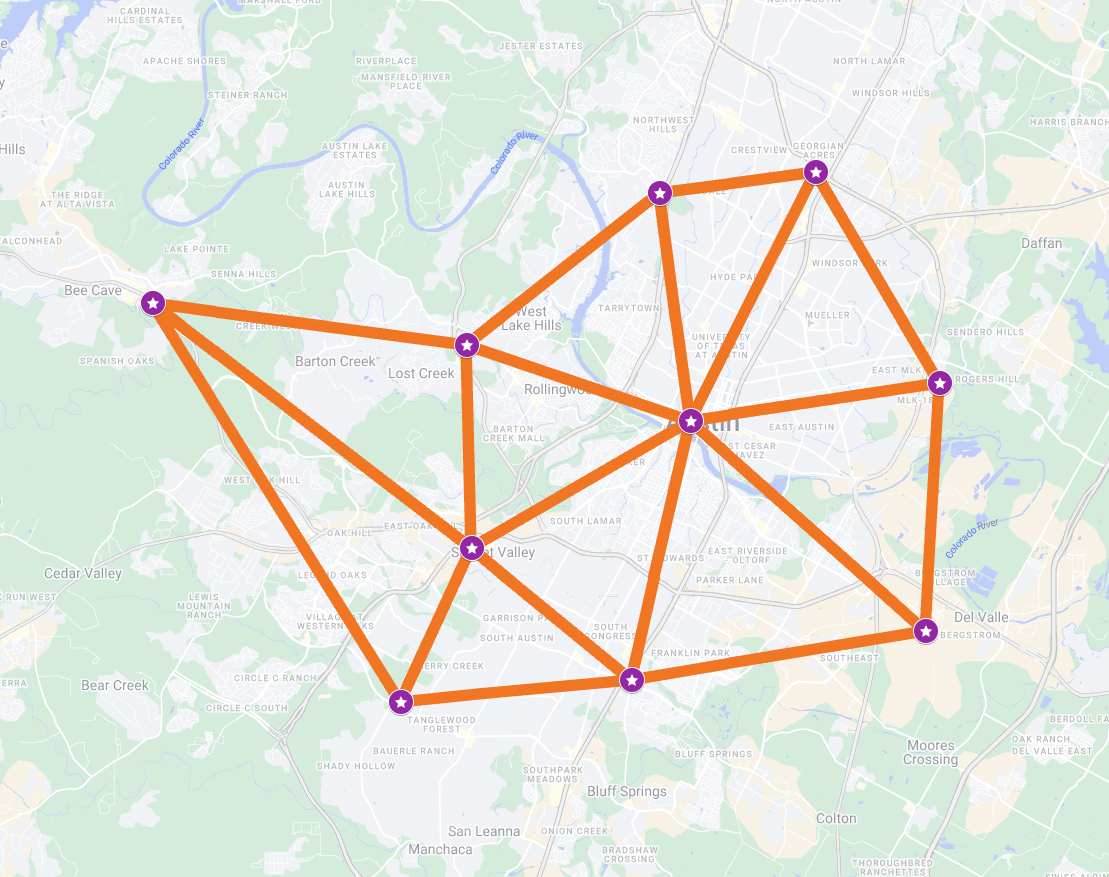}
     \caption{The South Austin UAM network in the case study}
    \label{fig:scenarioEnv}
\end{figure}

We choose the city of Austin as the case study to demonstrate our proposed RL approach for separated and quiet UAM traffic management. In this case study, we adopt a fraction of the Austin UAM network design in~\cite{gao2024noise} and form a South Austin UAM network that contains 10 destinations (vertiports), 19 unidirectional links (38 directional links) in each layer, and 28 origin-destination (O-D) pairs. \cref{fig:scenarioEnv} shows the structure of the South Austin UAM network. In addition, aircraft can operate at five different altitudes, ranging from 1,000 feet to 3,000 feet, with each altitude separated by 500-foot intervals.

We implement a simulation scenario involving 136 aircraft with similar performance to Eurocopter EC135 flying over the UAM network. At the beginning of each simulation, each aircraft is assigned a destination to which it must navigate. Each aircraft considers other aircraft within a \textit{communication distance} from its position when detecting intruders. In this scenario, we set the communication distance (\(d_{\text{comm}}\)) to 2.5 km. When determining safe separation, each aircraft only accounts for others on the same flight route or those on intersecting routes. The minimum separation distance (\(d_{\text{LOS}}\)) is set to 150 m. LOS events occur whenever two aircraft come within \(d_{\text{LOS}}\) of each other, measured as the Euclidean distance between them. Our objectives are to enforce safe separation between aircraft and minimize noise impact. This scenario remains consistent for both the training and testing of our RL models. During testing, we record metrics such as the number of LOS events and the cumulative noise increase over ambient levels. In addition,  to account for the varying ambient noise levels in the urban communities, the environment is partitioned into different noise zones, with each link and vertiport having its own ambient noise level. We calculate the cumulative noise increase for each zone with the cumulative noise of UAM operations and the ambient noise level, as described in \cref{eq:cumulative}. A lower number of LOS events indicates better safety performance, while a reduced cumulative noise increase signifies a decrease in noise impact.

\subsection{Simulation}

We train our reinforcement learning models using the BlueSky Air Traffic Simulator \cite{Simulator}, an open-source platform designed specifically for air traffic management research. 
With its open-data approach, BlueSky allows for the creation of custom, detailed scenarios using straightforward text-based commands, making it both accessible and flexible. 
Its modular design supports extensive customization, enabling researchers to incorporate advanced simulation elements, such as conflict detection and resolution algorithms, without relying on proprietary data or specialized hardware. 
We utilize BlueSky to simulate UAM scenarios in the South Austin network.
We define our simulation setup with a text-based scenario file that instantiates all aircraft traversing the network, specifying individual routes and takeoff times for each aircraft. 
Through repeated runs of this scenario, we train our reinforcement learning model to optimize noise and vertical separation objectives. 
BlueSky’s traffic module provides us with real-time aircraft states at each simulation time step, while its command stack allows for dynamic altitude adjustments, enabling our model to interact with and influence the simulation directly.

\section{Results}

We conducted experiments to evaluate the performance of various RL models across 100 episodes in a controlled testing scenario, which is the same as the training scenario. We train several policies with varying values of \(\rho \in [0, 1]\) in~\cref{eq:Rf}. Policies with \(\rho\) values close to 0 prioritize vertical separation between aircraft, while those with \(\rho\) values near 1 focus on mitigating noise impact. All policies are trained for 10000 iterations within the same scenario. The primary objective was to investigate how different reward function parameters affect aircraft behavior concerning altitude selection, noise impact, and safety performance\footnote{Code for the simulation testing environment is provided at~\url{https://github.com/suryakmurthy/UAM_noise_aware_agent}}.

\subsection{Altitude Selection and Noise Impact}

\begin{figure}[hbt!]
    \centering
    \begin{subfigure}[b]{0.18\textwidth}  
        \centering
        \scalebox{.91}{
\begin{tikzpicture}

\definecolor{chocolate197870}{RGB}{197,87,0}
\definecolor{darkslategray38}{RGB}{38,38,38}
\definecolor{lavender234234242}{RGB}{234,234,242}

\begin{axis}[
axis background/.style={fill=lavender234234242},
tick align=outside,
width = 4cm,
height = 4cm,
label style = {font=\footnotesize,},
ylabel style = {yshift=-9pt},
xlabel=\textcolor{darkslategray38}{Altitude Levels ($10^3$ ft)},
xmajorgrids,
xmajorticks=true,
xmin=-0.64, xmax=4.64,
xtick style={color=darkslategray38},
xtick={0,1,2,3,4},
xticklabels={1,1.5,2,2.5,3},
ymajorgrids,
ymajorticks=true,
ymin=0, ymax=1.05,
ytick style={color=darkslategray38}
]
\draw[draw=white,fill=chocolate197870,semithick] (axis cs:-0.4,0) rectangle (axis cs:0.4,0.380143202904397);
\draw[draw=white,fill=chocolate197870,semithick] (axis cs:0.6,0) rectangle (axis cs:1.4,0.174768051633723);
\draw[draw=white,fill=chocolate197870,semithick] (axis cs:1.6,0) rectangle (axis cs:2.4,0.107704719645018);
\draw[draw=white,fill=chocolate197870,semithick] (axis cs:2.6,0) rectangle (axis cs:3.4,0.142749092375958);
\draw[draw=white,fill=chocolate197870,semithick] (axis cs:3.6,0) rectangle (axis cs:4.4,0.194634933440904);
\end{axis}

\end{tikzpicture}}
        \caption{$\rho = 0$}
        \label{fig:cn-plot-0}
    \end{subfigure}
    \hfill
    \begin{subfigure}[b]{0.18\textwidth}
        \centering
        \scalebox{.91}{
\begin{tikzpicture}

\definecolor{chocolate197870}{RGB}{197,87,0}
\definecolor{darkslategray38}{RGB}{38,38,38}
\definecolor{lavender234234242}{RGB}{234,234,242}

\begin{axis}[
axis background/.style={fill=lavender234234242},
tick align=outside,
width = 4cm,
height = 4cm,
label style = {font=\footnotesize,},
ylabel style = {yshift=-7pt},
xlabel=\textcolor{darkslategray38}{Altitude Levels ($10^3$ ft)},
xmajorgrids,
xmajorticks=true,
xmin=-0.64, xmax=4.64,
xtick style={color=darkslategray38},
xtick={0,1,2,3,4},
xticklabels={1,1.5,2,2.5,3},
ymajorgrids,
ymajorticks=true,
ymin=0, ymax=1.05,
ytick style={color=darkslategray38}
]
\draw[draw=white,fill=chocolate197870,semithick] (axis cs:-0.4,0) rectangle (axis cs:0.4,0.174482723731472);
\draw[draw=white,fill=chocolate197870,semithick] (axis cs:0.6,0) rectangle (axis cs:1.4,0.175494511053777);
\draw[draw=white,fill=chocolate197870,semithick] (axis cs:1.6,0) rectangle (axis cs:2.4,0.278848586027217);
\draw[draw=white,fill=chocolate197870,semithick] (axis cs:2.6,0) rectangle (axis cs:3.4,0.205139879597309);
\draw[draw=white,fill=chocolate197870,semithick] (axis cs:3.6,0) rectangle (axis cs:4.4,0.166034299590226);
\end{axis}

\end{tikzpicture}}
        \caption{$\rho = 0.1$}
        \label{fig:cn-plot-01}
    \end{subfigure}
    \hfill
    \begin{subfigure}[b]{0.18\textwidth}
        \centering
        \scalebox{.91}{
\begin{tikzpicture}

\definecolor{chocolate197870}{RGB}{197,87,0}
\definecolor{darkslategray38}{RGB}{38,38,38}
\definecolor{lavender234234242}{RGB}{234,234,242}

\begin{axis}[
axis background/.style={fill=lavender234234242},
tick align=outside,
width = 4cm,
height = 4cm,
label style = {font=\footnotesize,},
ylabel style = {yshift=-7pt},
xlabel=\textcolor{darkslategray38}{Altitude Levels ($10^3$ ft)},
xmajorgrids,
xmajorticks=true,
xmin=-0.64, xmax=4.64,
xtick style={color=darkslategray38},
xtick={0,1,2,3,4},
xticklabels={1,1.5,2,2.5,3},
ymajorgrids,
ymajorticks=true,
ymin=0, ymax=1.05,
ytick style={color=darkslategray38}
]
\draw[draw=white,fill=chocolate197870,semithick] (axis cs:-0.4,0) rectangle (axis cs:0.4,0.0866910866910867);
\draw[draw=white,fill=chocolate197870,semithick] (axis cs:0.6,0) rectangle (axis cs:1.4,0.0812474562474562);
\draw[draw=white,fill=chocolate197870,semithick] (axis cs:1.6,0) rectangle (axis cs:2.4,0.16961741961742);
\draw[draw=white,fill=chocolate197870,semithick] (axis cs:2.6,0) rectangle (axis cs:3.4,0.254273504273504);
\draw[draw=white,fill=chocolate197870,semithick] (axis cs:3.6,0) rectangle (axis cs:4.4,0.408170533170533);
\end{axis}

\end{tikzpicture}}
        \caption{$\rho = 0.2$}
        \label{fig:cn-plot-02}
    \end{subfigure}
    \hfill
    \begin{subfigure}[b]{0.18\textwidth}
        \centering
        \scalebox{.91}{
\begin{tikzpicture}

\definecolor{chocolate197870}{RGB}{197,87,0}
\definecolor{darkslategray38}{RGB}{38,38,38}
\definecolor{lavender234234242}{RGB}{234,234,242}

\begin{axis}[
axis background/.style={fill=lavender234234242},
tick align=outside,
width = 4cm,
height = 4cm,
label style = {font=\footnotesize,},
ylabel style = {yshift=-7pt},
xlabel=\textcolor{darkslategray38}{Altitude Levels ($10^3$ ft)},
xmajorgrids,
xmajorticks=true,
xmin=-0.64, xmax=4.64,
xtick style={color=darkslategray38},
xtick={0,1,2,3,4},
xticklabels={1,1.5,2,2.5,3},
ymajorgrids,
ymajorticks=true,
ymin=0, ymax=1.05,
ytick style={color=darkslategray38}
]
\draw[draw=white,fill=chocolate197870,semithick] (axis cs:-0.4,0) rectangle (axis cs:0.4,0.0707215021941014);
\draw[draw=white,fill=chocolate197870,semithick] (axis cs:0.6,0) rectangle (axis cs:1.4,0.0423002347178284);
\draw[draw=white,fill=chocolate197870,semithick] (axis cs:1.6,0) rectangle (axis cs:2.4,0.078273293193183);
\draw[draw=white,fill=chocolate197870,semithick] (axis cs:2.6,0) rectangle (axis cs:3.4,0.20206143484029);
\draw[draw=white,fill=chocolate197870,semithick] (axis cs:3.6,0) rectangle (axis cs:4.4,0.606643535054597);
\end{axis}

\end{tikzpicture}}
        \caption{$\rho = 0.3$}
        \label{fig:cn-plot-03}
    \end{subfigure}
    \hfill
    \begin{subfigure}[b]{0.18\textwidth}
        \centering
        \scalebox{.91}{
\begin{tikzpicture}

\definecolor{chocolate197870}{RGB}{197,87,0}
\definecolor{darkslategray38}{RGB}{38,38,38}
\definecolor{lavender234234242}{RGB}{234,234,242}

\begin{axis}[
axis background/.style={fill=lavender234234242},
tick align=outside,
width = 4cm,
height = 4cm,
label style = {font=\footnotesize,},
ylabel style = {yshift=-7pt},
xlabel=\textcolor{darkslategray38}{Altitude Levels ($10^3$ ft)},
xmajorgrids,
xmajorticks=true,
xmin=-0.64, xmax=4.64,
xtick style={color=darkslategray38},
xtick={0,1,2,3,4},
xticklabels={1,1.5,2,2.5,3},
ymajorgrids,
ymajorticks=true,
ymin=0, ymax=1.05,
ytick style={color=darkslategray38}
]
\draw[draw=white,fill=chocolate197870,semithick] (axis cs:-0.4,0) rectangle (axis cs:0.4,0.0704964828218983);
\draw[draw=white,fill=chocolate197870,semithick] (axis cs:0.6,0) rectangle (axis cs:1.4,0.0388418799062086);
\draw[draw=white,fill=chocolate197870,semithick] (axis cs:1.6,0) rectangle (axis cs:2.4,0.0897135283922928);
\draw[draw=white,fill=chocolate197870,semithick] (axis cs:2.6,0) rectangle (axis cs:3.4,0.291568967274952);
\draw[draw=white,fill=chocolate197870,semithick] (axis cs:3.6,0) rectangle (axis cs:4.4,0.509379141604649);
\end{axis}

\end{tikzpicture}}
        \caption{$\rho = 0.4$}
        \label{fig:cn-plot-04}
    \end{subfigure}
    \begin{subfigure}[b]{0.18\textwidth}
        \centering
        \scalebox{.91}{
\begin{tikzpicture}

\definecolor{chocolate197870}{RGB}{197,87,0}
\definecolor{darkslategray38}{RGB}{38,38,38}
\definecolor{lavender234234242}{RGB}{234,234,242}

\begin{axis}[
axis background/.style={fill=lavender234234242},
tick align=outside,
width = 4cm,
height = 4cm,
label style = {font=\footnotesize,},
ylabel style = {yshift=-9pt},
xlabel=\textcolor{darkslategray38}{Altitude Levels ($10^3$ ft)},
xmajorgrids,
xmajorticks=true,
xmin=-0.64, xmax=4.64,
xtick style={color=darkslategray38},
xtick={0,1,2,3,4},
xticklabels={1,1.5,2,2.5,3},
ymajorgrids,
ymajorticks=true,
ymin=0, ymax=1.05,
ytick style={color=darkslategray38}
]
\draw[draw=white,fill=chocolate197870,semithick] (axis cs:-0.4,0) rectangle (axis cs:0.4,0.0705360330494211);
\draw[draw=white,fill=chocolate197870,semithick] (axis cs:0.6,0) rectangle (axis cs:1.4,0.0346304891110318);
\draw[draw=white,fill=chocolate197870,semithick] (axis cs:1.6,0) rectangle (axis cs:2.4,0.0536543071352068);
\draw[draw=white,fill=chocolate197870,semithick] (axis cs:2.6,0) rectangle (axis cs:3.4,0.292701586168205);
\draw[draw=white,fill=chocolate197870,semithick] (axis cs:3.6,0) rectangle (axis cs:4.4,0.548477584536135);
\end{axis}

\end{tikzpicture}}
        \caption{$\rho = 0.5$}
        \label{fig:cn-plot-05}
    \end{subfigure}
    \hfill
    \begin{subfigure}[b]{0.18\textwidth}
        \centering
        \scalebox{.91}{
\begin{tikzpicture}

\definecolor{chocolate197870}{RGB}{197,87,0}
\definecolor{darkslategray38}{RGB}{38,38,38}
\definecolor{lavender234234242}{RGB}{234,234,242}

\begin{axis}[
axis background/.style={fill=lavender234234242},
tick align=outside,
width = 4cm,
height = 4cm,
label style = {font=\footnotesize,},
ylabel style = {yshift=-9pt},
xlabel=\textcolor{darkslategray38}{Altitude Levels ($10^3$ ft)},
xmajorgrids,
xmajorticks=true,
xmin=-0.64, xmax=4.64,
xtick style={color=darkslategray38},
xtick={0,1,2,3,4},
xticklabels={1,1.5,2,2.5,3},
ymajorgrids,
ymajorticks=true,
ymin=0, ymax=1.05,
ytick style={color=darkslategray38}
]
\draw[draw=white,fill=chocolate197870,semithick] (axis cs:-0.4,0) rectangle (axis cs:0.4,0.073351381723451);
\draw[draw=white,fill=chocolate197870,semithick] (axis cs:0.6,0) rectangle (axis cs:1.4,0.0416815651019053);
\draw[draw=white,fill=chocolate197870,semithick] (axis cs:1.6,0) rectangle (axis cs:2.4,0.0504673851969147);
\draw[draw=white,fill=chocolate197870,semithick] (axis cs:2.6,0) rectangle (axis cs:3.4,0.0876538795525361);
\draw[draw=white,fill=chocolate197870,semithick] (axis cs:3.6,0) rectangle (axis cs:4.4,0.746845788425193);
\end{axis}

\end{tikzpicture}}
        \caption{$\rho = 0.6$}
        \label{fig:cn-plot-06}
    \end{subfigure}
    \hfill
    \begin{subfigure}[b]{0.18\textwidth}
        \centering
        \scalebox{.91}{
\begin{tikzpicture}

\definecolor{chocolate197870}{RGB}{197,87,0}
\definecolor{darkslategray38}{RGB}{38,38,38}
\definecolor{lavender234234242}{RGB}{234,234,242}

\begin{axis}[
axis background/.style={fill=lavender234234242},
tick align=outside,
width = 4cm,
height = 4cm,
label style = {font=\footnotesize,},
ylabel style = {yshift=-9pt},
xlabel=\textcolor{darkslategray38}{Altitude Levels ($10^3$ ft)},
xmajorgrids,
xmajorticks=true,
xmin=-0.64, xmax=4.64,
xtick style={color=darkslategray38},
xtick={0,1,2,3,4},
xticklabels={1,1.5,2,2.5,3},
ymajorgrids,
ymajorticks=true,
ymin=0, ymax=1.05,
ytick style={color=darkslategray38}
]
\draw[draw=white,fill=chocolate197870,semithick] (axis cs:-0.4,0) rectangle (axis cs:0.4,0.0706874520828009);
\draw[draw=white,fill=chocolate197870,semithick] (axis cs:0.6,0) rectangle (axis cs:1.4,0.0347048300536673);
\draw[draw=white,fill=chocolate197870,semithick] (axis cs:1.6,0) rectangle (axis cs:2.4,0.0347048300536673);
\draw[draw=white,fill=chocolate197870,semithick] (axis cs:2.6,0) rectangle (axis cs:3.4,0.0347048300536673);
\draw[draw=white,fill=chocolate197870,semithick] (axis cs:3.6,0) rectangle (axis cs:4.4,0.825198057756197);
\end{axis}

\end{tikzpicture}}
        \caption{$\rho = 0.7$}
        \label{fig:cn-plot-07}
    \end{subfigure}
    \hfill
    \begin{subfigure}[b]{0.18\textwidth}
        \centering
        \scalebox{.91}{
\begin{tikzpicture}

\definecolor{chocolate197870}{RGB}{197,87,0}
\definecolor{darkslategray38}{RGB}{38,38,38}
\definecolor{lavender234234242}{RGB}{234,234,242}

\begin{axis}[
axis background/.style={fill=lavender234234242},
tick align=outside,
width = 4cm,
height = 4cm,
label style = {font=\footnotesize,},
ylabel style = {yshift=-9pt},
xlabel=\textcolor{darkslategray38}{Altitude Levels ($10^3$ ft)},
xmajorgrids,
xmajorticks=true,
xmin=-0.64, xmax=4.64,
xtick style={color=darkslategray38},
xtick={0,1,2,3,4},
xticklabels={1,1.5,2,2.5,3},
ymajorgrids,
ymajorticks=true,
ymin=0, ymax=1.05,
ytick style={color=darkslategray38}
]
\draw[draw=white,fill=chocolate197870,semithick] (axis cs:-0.4,0) rectangle (axis cs:0.4,0.0706874520828009);
\draw[draw=white,fill=chocolate197870,semithick] (axis cs:0.6,0) rectangle (axis cs:1.4,0.0347048300536673);
\draw[draw=white,fill=chocolate197870,semithick] (axis cs:1.6,0) rectangle (axis cs:2.4,0.0347048300536673);
\draw[draw=white,fill=chocolate197870,semithick] (axis cs:2.6,0) rectangle (axis cs:3.4,0.0347048300536673);
\draw[draw=white,fill=chocolate197870,semithick] (axis cs:3.6,0) rectangle (axis cs:4.4,0.825198057756197);
\end{axis}

\end{tikzpicture}}
        \caption{$\rho = 0.8$}
        \label{fig:cn-plot-08}
    \end{subfigure}
    \hfill
    \begin{subfigure}[b]{0.18\textwidth}
        \centering
        \scalebox{.91}{
\begin{tikzpicture}

\definecolor{chocolate197870}{RGB}{197,87,0}
\definecolor{darkslategray38}{RGB}{38,38,38}
\definecolor{lavender234234242}{RGB}{234,234,242}

\begin{axis}[
axis background/.style={fill=lavender234234242},
tick align=outside,
width = 4cm,
height = 4cm,
label style = {font=\footnotesize,},
ylabel style = {yshift=-9pt},
xlabel=\textcolor{darkslategray38}{Altitude Levels ($10^3$ ft)},
xmajorgrids,
xmajorticks=true,
xmin=-0.64, xmax=4.64,
xtick style={color=darkslategray38},
xtick={0,1,2,3,4},
xticklabels={1,1.5,2,2.5,3},
ymajorgrids,
ymajorticks=true,
ymin=0, ymax=1.05,
ytick style={color=darkslategray38}
]
\draw[draw=white,fill=chocolate197870,semithick] (axis cs:-0.4,0) rectangle (axis cs:0.4,0.0706874520828009);
\draw[draw=white,fill=chocolate197870,semithick] (axis cs:0.6,0) rectangle (axis cs:1.4,0.0347048300536673);
\draw[draw=white,fill=chocolate197870,semithick] (axis cs:1.6,0) rectangle (axis cs:2.4,0.0347048300536673);
\draw[draw=white,fill=chocolate197870,semithick] (axis cs:2.6,0) rectangle (axis cs:3.4,0.0347048300536673);
\draw[draw=white,fill=chocolate197870,semithick] (axis cs:3.6,0) rectangle (axis cs:4.4,0.825198057756197);
\end{axis}

\end{tikzpicture}}
        \caption{$\rho = 0.9$}
        \label{fig:cn-plot-09}
    \end{subfigure}

    \caption{Air traffic altitude distribution plots with varying tradeoff hyperparameter values from 0 to 0.9. Increasing emphasis on noise leads to increased traffic congestion at the highest altitudes.}
    \label{fig:cn-plots-grid}
\end{figure}
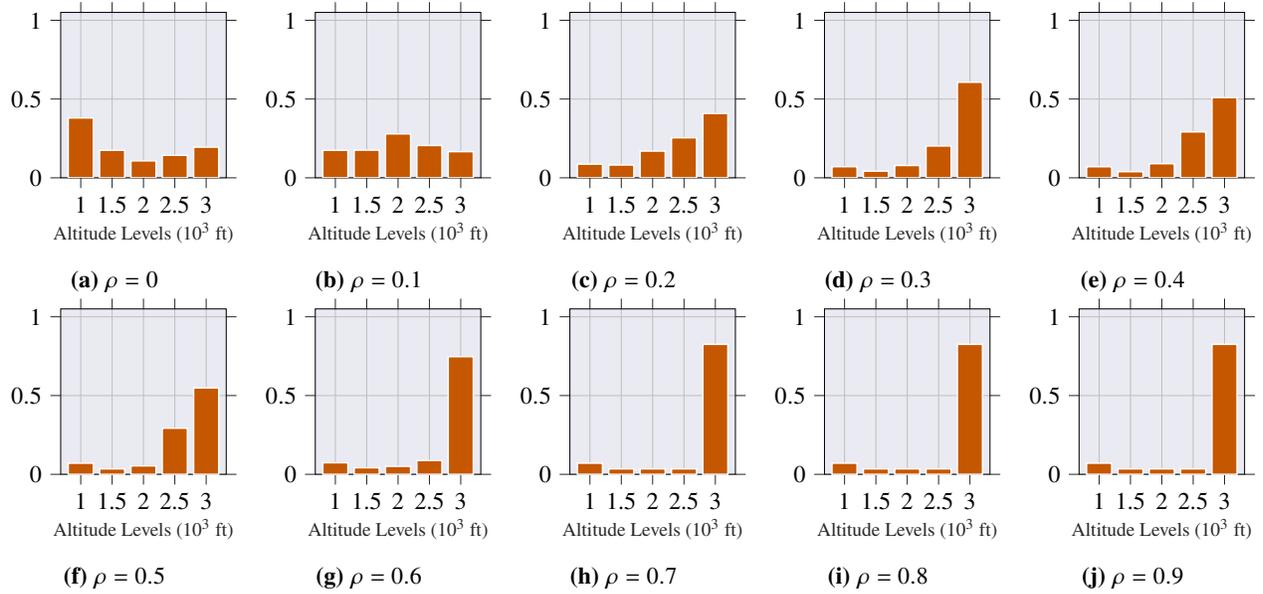

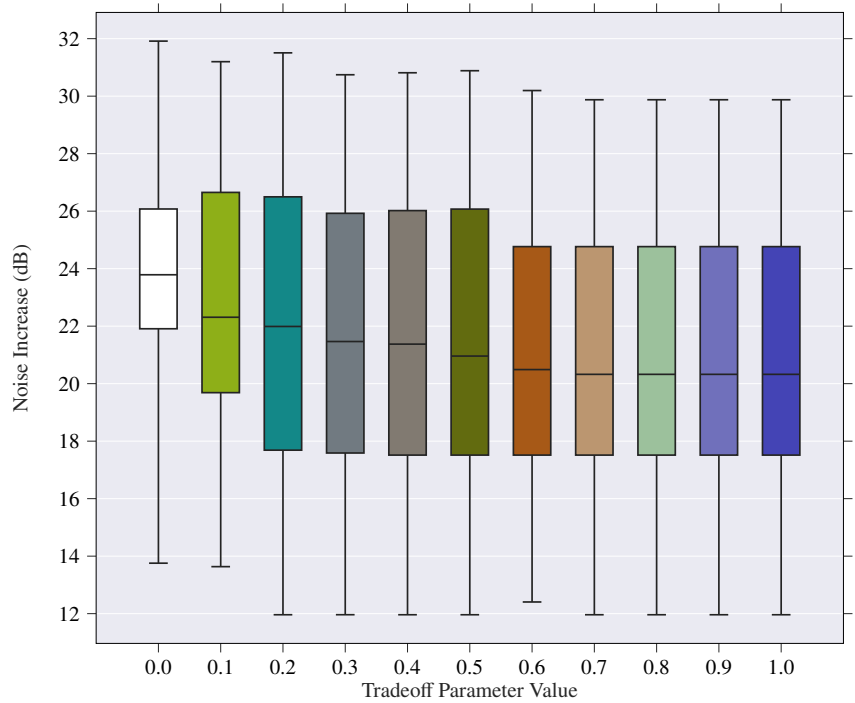
\begin{figure}[hbt!]
    \centering
    \scalebox{.8}{
\begin{tikzpicture}

\definecolor{darkcyan19136136}{RGB}{19,136,136}
\definecolor{darkolivegreen9810715}{RGB}{98,107,15}
\definecolor{darkseagreen156193156}{RGB}{156,193,156}
\definecolor{darkslateblue6868181}{RGB}{68,68,181}
\definecolor{darkslategray36}{RGB}{36,36,36}
\definecolor{darkslategray38}{RGB}{38,38,38}
\definecolor{gray113122129}{RGB}{113,122,129}
\definecolor{gray129122113}{RGB}{129,122,113}
\definecolor{lavender234234242}{RGB}{234,234,242}
\definecolor{rosybrown187150112}{RGB}{187,150,112}
\definecolor{sienna1678923}{RGB}{167,89,23}
\definecolor{slateblue112112187}{RGB}{112,112,187}
\definecolor{yellowgreen14217524}{RGB}{142,175,24}

\begin{axis}[
    width=14cm,  
    axis background/.style={fill=lavender234234242},
    tick align=outside,
    x grid style={white},
    xlabel=\textcolor{darkslategray38}{Tradeoff Parameter Value},
    xmajorticks=true,  
    xmin=-1, xmax=11,  
    xtick style={color=darkslategray38},
    xtick={0,1,2,3,4,5,6,7,8,9,10},
    xticklabels={0.0,0.1,0.2,0.3,0.4,0.5,0.6,0.7,0.8,0.9,1.0},
    y grid style={white},
    ylabel=\textcolor{darkslategray38}{Noise Increase (dB)},
    ymajorgrids,
    ymajorticks=true,  
    ymin=10.9649539449411, ymax=32.9123571249206,
    ytick style={color=darkslategray38}
]

\path [draw=darkslategray36, fill=white, thick]
(axis cs:-0.3,21.9092939368943)
--(axis cs:0.3,21.9092939368943)
--(axis cs:0.3,26.0753513165004)
--(axis cs:-0.3,26.0753513165004)
--(axis cs:-0.3,21.9092939368943)
--cycle;
\addplot [thick, darkslategray36]
table {%
0 21.9092939368943
0 13.7555454816011
};
\addplot [thick, darkslategray36]
table {%
0 26.0753513165004
0 31.914747889467
};
\addplot [thick, darkslategray36]
table {%
-0.15 13.7555454816011
0.15 13.7555454816011
};
\addplot [thick, darkslategray36]
table {%
-0.15 31.914747889467
0.15 31.914747889467
};
\path [draw=darkslategray36, fill=yellowgreen14217524, thick]
(axis cs:0.7,19.6867285475727)
--(axis cs:1.3,19.6867285475727)
--(axis cs:1.3,26.6519498415858)
--(axis cs:0.7,26.6519498415858)
--(axis cs:0.7,19.6867285475727)
--cycle;
\addplot [thick, darkslategray36]
table {%
1 19.6867285475727
1 13.634940592712
};
\addplot [thick, darkslategray36]
table {%
1 26.6519498415858
1 31.1990327215267
};
\addplot [thick, darkslategray36]
table {%
0.85 13.634940592712
1.15 13.634940592712
};
\addplot [thick, darkslategray36]
table {%
0.85 31.1990327215267
1.15 31.1990327215267
};
\path [draw=darkslategray36, fill=darkcyan19136136, thick]
(axis cs:1.7,17.6865629435354)
--(axis cs:2.3,17.6865629435354)
--(axis cs:2.3,26.4980334873487)
--(axis cs:1.7,26.4980334873487)
--(axis cs:1.7,17.6865629435354)
--cycle;
\addplot [thick, darkslategray36]
table {%
2 17.6865629435354
2 11.9625631803947
};
\addplot [thick, darkslategray36]
table {%
2 26.4980334873487
2 31.5065191181259
};
\addplot [thick, darkslategray36]
table {%
1.85 11.9625631803947
2.15 11.9625631803947
};
\addplot [thick, darkslategray36]
table {%
1.85 31.5065191181259
2.15 31.5065191181259
};
\path [draw=darkslategray36, fill=gray113122129, thick]
(axis cs:2.7,17.5875093720813)
--(axis cs:3.3,17.5875093720813)
--(axis cs:3.3,25.9247938489723)
--(axis cs:2.7,25.9247938489723)
--(axis cs:2.7,17.5875093720813)
--cycle;
\addplot [thick, darkslategray36]
table {%
3 17.5875093720813
3 11.9635343140649
};
\addplot [thick, darkslategray36]
table {%
3 25.9247938489723
3 30.7445348702963
};
\addplot [thick, darkslategray36]
table {%
2.85 11.9635343140649
3.15 11.9635343140649
};
\addplot [thick, darkslategray36]
table {%
2.85 30.7445348702963
3.15 30.7445348702963
};
\path [draw=darkslategray36, fill=gray129122113, thick]
(axis cs:3.7,17.5148391893344)
--(axis cs:4.3,17.5148391893344)
--(axis cs:4.3,26.0195115012575)
--(axis cs:3.7,26.0195115012575)
--(axis cs:3.7,17.5148391893344)
--cycle;
\addplot [thick, darkslategray36]
table {%
4 17.5148391893344
4 11.9625631803947
};
\addplot [thick, darkslategray36]
table {%
4 26.0195115012575
4 30.8136680669647
};
\addplot [thick, darkslategray36]
table {%
3.85 11.9625631803947
4.15 11.9625631803947
};
\addplot [thick, darkslategray36]
table {%
3.85 30.8136680669647
4.15 30.8136680669647
};
\path [draw=darkslategray36, fill=darkolivegreen9810715, thick]
(axis cs:4.7,17.5147556589798)
--(axis cs:5.3,17.5147556589798)
--(axis cs:5.3,26.0713419644216)
--(axis cs:4.7,26.0713419644216)
--(axis cs:4.7,17.5147556589798)
--cycle;
\addplot [thick, darkslategray36]
table {%
5 17.5147556589798
5 11.9625631803947
};
\addplot [thick, darkslategray36]
table {%
5 26.0713419644216
5 30.8848699852602
};
\addplot [thick, darkslategray36]
table {%
4.85 11.9625631803947
5.15 11.9625631803947
};
\addplot [thick, darkslategray36]
table {%
4.85 30.8848699852602
5.15 30.8848699852602
};
\path [draw=darkslategray36, fill=sienna1678923, thick]
(axis cs:5.7,17.5147556589798)
--(axis cs:6.3,17.5147556589798)
--(axis cs:6.3,24.7665720350297)
--(axis cs:5.7,24.7665720350297)
--(axis cs:5.7,17.5147556589798)
--cycle;
\addplot [thick, darkslategray36]
table {%
6 17.5147556589798
6 12.4058901759531
};
\addplot [thick, darkslategray36]
table {%
6 24.7665720350297
6 30.1952327097984
};
\addplot [thick, darkslategray36]
table {%
5.85 12.4058901759531
6.15 12.4058901759531
};
\addplot [thick, darkslategray36]
table {%
5.85 30.1952327097984
6.15 30.1952327097984
};
\path [draw=darkslategray36, fill=rosybrown187150112, thick]
(axis cs:6.7,17.5147556589798)
--(axis cs:7.3,17.5147556589798)
--(axis cs:7.3,24.7661168970678)
--(axis cs:6.7,24.7661168970678)
--(axis cs:6.7,17.5147556589798)
--cycle;
\addplot [thick, darkslategray36]
table {%
7 17.5147556589798
7 11.9625631803947
};
\addplot [thick, darkslategray36]
table {%
7 24.7661168970678
7 29.8747841129995
};
\addplot [thick, darkslategray36]
table {%
6.85 11.9625631803947
7.15 11.9625631803947
};
\addplot [thick, darkslategray36]
table {%
6.85 29.8747841129995
7.15 29.8747841129995
};
\path [draw=darkslategray36, fill=darkseagreen156193156, thick]
(axis cs:7.7,17.5147556589798)
--(axis cs:8.3,17.5147556589798)
--(axis cs:8.3,24.7661168970678)
--(axis cs:7.7,24.7661168970678)
--(axis cs:7.7,17.5147556589798)
--cycle;
\addplot [thick, darkslategray36]
table {%
8 17.5147556589798
8 11.9625631803947
};
\addplot [thick, darkslategray36]
table {%
8 24.7661168970678
8 29.8747841129995
};
\addplot [thick, darkslategray36]
table {%
7.85 11.9625631803947
8.15 11.9625631803947
};
\addplot [thick, darkslategray36]
table {%
7.85 29.8747841129995
8.15 29.8747841129995
};
\path [draw=darkslategray36, fill=slateblue112112187, thick]
(axis cs:8.7,17.5147556589798)
--(axis cs:9.3,17.5147556589798)
--(axis cs:9.3,24.7665826809562)
--(axis cs:8.7,24.7665826809562)
--(axis cs:8.7,17.5147556589798)
--cycle;
\addplot [thick, darkslategray36]
table {%
9 17.5147556589798
9 11.9625631803947
};
\addplot [thick, darkslategray36]
table {%
9 24.7665826809562
9 29.8747841129995
};
\addplot [thick, darkslategray36]
table {%
8.85 11.9625631803947
9.15 11.9625631803947
};
\addplot [thick, darkslategray36]
table {%
8.85 29.8747841129995
9.15 29.8747841129995
};
\path [draw=darkslategray36, fill=darkslateblue6868181, thick]
(axis cs:9.7,17.5147556589798)
--(axis cs:10.3,17.5147556589798)
--(axis cs:10.3,24.7661168970678)
--(axis cs:9.7,24.7661168970678)
--(axis cs:9.7,17.5147556589798)
--cycle;
\addplot [thick, darkslategray36]
table {%
10 17.5147556589798
10 11.9625631803947
};
\addplot [thick, darkslategray36]
table {%
10 24.7661168970678
10 29.8747841129995
};
\addplot [thick, darkslategray36]
table {%
9.85 11.9625631803947
10.15 11.9625631803947
};
\addplot [thick, darkslategray36]
table {%
9.85 29.8747841129995
10.15 29.8747841129995
};
\addplot [thick, darkslategray36]
table {%
-0.3 23.7874954279433
0.3 23.7874954279433
};
\addplot [thick, darkslategray36]
table {%
0.7 22.3097108906162
1.3 22.3097108906162
};
\addplot [thick, darkslategray36]
table {%
1.7 21.9889140717458
2.3 21.9889140717458
};
\addplot [thick, darkslategray36]
table {%
2.7 21.4653817064495
3.3 21.4653817064495
};
\addplot [thick, darkslategray36]
table {%
3.7 21.3750674199901
4.3 21.3750674199901
};
\addplot [thick, darkslategray36]
table {%
4.7 20.9596024120096
5.3 20.9596024120096
};
\addplot [thick, darkslategray36]
table {%
5.7 20.4895032570277
6.3 20.4895032570277
};
\addplot [thick, darkslategray36]
table {%
6.7 20.3238885653638
7.3 20.3238885653638
};
\addplot [thick, darkslategray36]
table {%
7.7 20.3238885653638
8.3 20.3238885653638
};
\addplot [thick, darkslategray36]
table {%
8.7 20.3238885653638
9.3 20.3238885653638
};
\addplot [thick, darkslategray36]
table {%
9.7 20.3238885653638
10.3 20.3238885653638
};
\end{axis}

\end{tikzpicture}}
    \caption{Cumulative noise increase over ambient levels per experiment group. We observe that increasing emphasis on noise in the reward function leads to a decrease in noise impact.}
    \label{fig:noise_comparison}
\end{figure}

The altitude selection result is shown in \cref{fig:cn-plots-grid}. We can observe that the increase in the tradeoff parameter \(\rho\) leads the trained policy to favor the highest possible altitude. This behavior directly correlates with a reduction in cumulative noise impact, as demonstrated in \cref{fig:noise_comparison}. Higher altitudes typically lessen the noise impact on the ground, as flight trajectories are farther from populated areas, aligning with the noise reduction objective. Specifically, we observe that when \(\rho\) reaches values closer to 1, aircraft consistently prefer to operate at the maximum altitude available within the scenario. For values of \(\rho \geq 0.7\), the noise cost dominates the vertical separation cost, leading to the uniform behavior where aircraft fly at the highest possible altitude at all times. This trend in altitude preference, while beneficial for noise reduction, has implications for air traffic congestion and safety, as will be discussed next. Conversely, when $\rho$ approaches 0, the RL model promotes greater diversity in altitude selection and a more balanced distribution of traffic across all five altitude levels.

\vspace*{1cm}

\subsection{Safe Separation and Traffic Distribution}

\begin{figure}[hbt!]
    \centering
    \begin{subfigure}[b]{0.48\textwidth}
        \centering
        \scalebox{.93}{\begin{tikzpicture}
\definecolor{darkslategray}{RGB}{47,79,79}  
\definecolor{steelblue}{RGB}{70,130,180}    
\definecolor{lavender}{RGB}{230,230,250}     

\begin{axis}[
    axis background/.style={fill=lavender},
    axis line style={white},
    tick align=outside,
    xlabel={Median Noise Increase},
    ylabel={Percentage of Aircraft at Maximum Altitude},
    grid=both,
    major grid style={line width=.2pt,draw=gray!50},
    minor grid style={line width=.1pt,draw=gray!20},
    xmajorgrids,
    ymajorgrids,
    xmin=20, xmax=24,
    ymin=0, ymax=1,
    xtick style={color=darkslategray!80},
    ytick style={color=darkslategray!80},
]

\addplot [
    line width=1pt, 
    color=steelblue,
    mark=*,
    mark size=4,
    mark options={solid}
]
table {
    23.7874954279433 0.199402501459243
    22.3097108906162 0.20697964622732
    21.9889140717458 0.432016569514541
    21.4653817064495 0.569089390142022
    21.3750674199901 0.536224144809503
    20.9596024120096 0.542909956986954
    20.4895032570277 0.746314748083776
    20.3238885653638 0.819584750494949
    20.3238885653638 0.819584750494949
    20.3238885653638 0.81958385989849
    20.3238885653638 0.819581124495079
};
\node[anchor=west, text=darkslategray, font=\small] at (35,85) {$\rho = 0.7 - 1$}; 
\node[anchor=west, text=darkslategray, font=\small] at (53,77) {$\rho = 0.6$}; 
\node[anchor=west, text=darkslategray, font=\small] at (85,63) {$\rho = 0.5$}; 
\node[anchor=west, text=darkslategray, font=\small] at (137,51) {$\rho = 0.4$}; 
\node[anchor=west, text=darkslategray, font=\small] at (150,60) {$\rho = 0.3$}; 
\node[anchor=west, text=darkslategray, font=\small] at (203,43) {$\rho = 0.2$}; 
\node[anchor=west, text=darkslategray, font=\small] at (223,27) {$\rho = 0.1$}; 
\node[anchor=west, text=darkslategray, font=\small] at (335,27) {$\rho = 0$}; 
\end{axis}

\end{tikzpicture}}
        \caption{The tradeoff between traffic congestion and noise increase. Reducing noise impact leads to increased traffic congestion at the highest altitude level.}
        \label{fig:tradeoff-congestion-noise}
    \end{subfigure}
    \hfill
    \begin{subfigure}[b]{0.48\textwidth}
        \centering
        \scalebox{.94}{
\begin{tikzpicture}

\definecolor{darkslategray38}{RGB}{38,38,38}
\definecolor{lavender234234242}{RGB}{234,234,242}
\definecolor{steelblue76114176}{RGB}{76,114,176}

\begin{axis}[
axis background/.style={fill=lavender234234242},
axis line style={white},
tick align=outside,
x grid style={white},
xlabel=\textcolor{darkslategray38}{Median Noise Increase},
xmajorgrids,
xmin=20.15, xmax=23.96,
xtick style={color=darkslategray38},
y grid style={white},
ylabel=\textcolor{darkslategray38}{Average LOS Occurrences},
ymajorgrids,
ymin=2.38, ymax=21.62,
ytick style={color=darkslategray38}
]

\draw[blue, thick] (axis cs:23.787,4.357) -- (axis cs:23.787,8.696);
\draw[blue, thick] (axis cs:22.310,3.261) -- (axis cs:22.310,9.475);
\draw[blue, thick] (axis cs:21.989,8.687) -- (axis cs:21.989,16.014);
\draw[blue, thick] (axis cs:21.465,10.020) -- (axis cs:21.465,17.138);
\draw[blue, thick] (axis cs:21.375,10.733) -- (axis cs:21.375,16.215);
\draw[blue, thick] (axis cs:20.960,13.562) -- (axis cs:20.960,16.929);
\draw[blue, thick] (axis cs:20.490,13.463) -- (axis cs:20.490,20.748);
\draw[blue, thick] (axis cs:20.324,16) -- (axis cs:20.324,16);

\addplot [line width=0.9pt, steelblue76114176, mark=*, mark size=4.5, mark options={solid}]
coordinates {
(23.787, 6.526)
(22.310, 6.368)
(21.989, 12.351)
(21.465, 13.579)
(21.375, 13.474)
(20.960, 15.246)
(20.490, 17.105)
(20.324, 16)
};

\addplot [line width=0.9pt, blue, mark=-, mark size=5, mark options={solid}, only marks]
table {
23.787 4.357
22.310 3.261
21.989 8.687
21.465 10.020
21.375 10.733
20.960 13.562
20.490 13.463
};

\addplot [line width=0.9pt, blue, mark=-, mark size=5, mark options={solid}, only marks]
table {
23.787 8.696
22.310 9.475
21.989 16.014
21.465 17.138
21.375 16.215
20.960 16.929
20.490 20.748
};

\draw (axis cs:23.387,6.968) node[scale=0.55, anchor=west, text=darkslategray38] {$\rho = 0$};
\draw (axis cs:22.410,6.968) node[scale=0.55, anchor=west, text=darkslategray38] {$\rho = 0.1$};
\draw (axis cs:22.089,12.351) node[scale=0.55, anchor=west, text=darkslategray38] {$\rho = 0.2$};
\draw (axis cs:21.565,14.5) node[scale=0.55, anchor=west, text=darkslategray38] {$\rho = 0.3$};
\draw (axis cs:20.9,12.6) node[scale=0.55, anchor=west, text=darkslategray38] {$\rho = 0.4$};
\draw (axis cs:20.560,14.046) node[scale=0.55, anchor=west, text=darkslategray38] {$\rho = 0.5$};
\draw (axis cs:20.590,17.405) node[scale=0.55, anchor=west, text=darkslategray38] {$\rho = 0.6$};
\draw (axis cs:20.124,14.8) node[scale=0.55, anchor=west, text=darkslategray38] {$\rho=0.7$};

\end{axis}

\end{tikzpicture}}
        \caption{The tradeoff between noise increase and loss of separation events. Reducing noise impact leads to increased number of LOS events.}
        \label{fig:tradeoff-safety-noise}
    \end{subfigure}
    \caption{The tradeoff plots: (a) congestion vs noise increase, and (b) loss of separation vs noise increase.}
    \label{fig:combined-tradeoff-plots}
\end{figure}
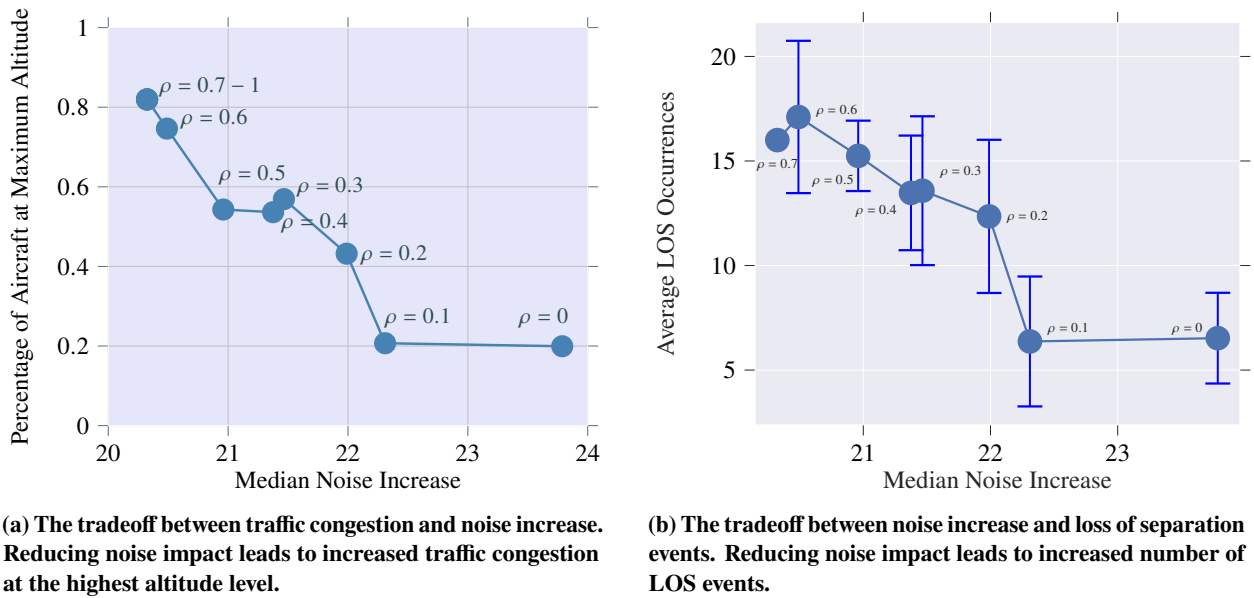

In \cref{fig:tradeoff-congestion-noise}, we explore the relationship between noise impact and vertical separation assurance. When \(\rho\) is close to 1, the majority of aircraft operate at the highest altitude, leading to a reduced noise impact. In the meantime, when the majority of aircraft operate at the maximum available altitude, congestion increases, along with the likelihood of LOS events. LOS events represent instances where aircraft breach the minimum safe separation distance, posing a risk to air safety. Thus, prioritizing noise reduction by increasing \(\rho\) could potentially lead to a compromise in safety, highlighting a critical tradeoff in UAM operations. As aircraft operate at different levels, there is a broader distribution of traffic across altitudes, reducing overall congestion at any single altitude level. This behavior, in turn, results in fewer LOS events, as shown in \cref{fig:tradeoff-safety-noise}. By spreading aircraft across multiple altitudes, the model effectively enforces a safe separation between aircraft, enhancing safety by reducing LOS risk. However, this strategy leads to increased cumulative noise impact, because more aircraft are operating closer to the ground. Hence, prioritizing safe vertical separation by setting \(\rho\) to lower values promotes safety but at the expense of increased noise impact to the ground. This is also undesirable for UAM systems operating in urban environments, where noise pollution poses a significant concern for local communities.

\subsection{Tradeoffs Between Noise, Safety, and Congestion}

Our findings suggest a tradeoff between real-time vertical separation assurance and noise mitigation. A solution focusing on minimizing noise tends to assign aircraft to higher altitudes, which leads to increased traffic congestion and a higher likelihood of LOS events. In contrast, a solution that prioritizes maintaining safe vertical separation leads to a more dispersed distribution across altitudes, thereby decreasing the occurrence of LOS events while increasing the noise impact on the ground. By maintaining vertical spacing between aircraft, emphasizing vertical separation not only reduces LOS events but also alleviates congestion at the highest altitude. This balance between noise and safety is essential for UAM system design, as different operational priorities can lead to markedly different outcomes in terms of environmental impact and air traffic efficiency. Our RL approach is an effective response to this tradeoff by making altitude adjustments based on local observations and reward function weighting. Future UAM systems will need to carefully consider this balance, or utilize additional methods to ensure safe separation between aircraft.

\section{Conclusion and Future Work}\label{sec:conclusion}

In this work, we explored the use of reinforcement learning for the real-time mitigation of UAM's noise impact.
We presented an RL model to achieve both vertical separation assurance and noise mitigation in air traffic management.
The results indicate a tradeoff between reducing traffic congestion and LOS occurrences at different altitude levels and the overall noise impact of the system.
Avenues for future work include augmenting our reinforcement learning model to account for other objectives. 
This includes adding a penalty for energy consumption associated with altitude adjustments.
In such cases, the reinforcement learning model would need to consider other tradeoffs, such as the noise reduction associated with the altitude adjustment versus the energy cost of making the adjustment.
In a second avenue for future work, one can consider alternative approaches to real-time noise mitigation and safety assurance.
Other tactical separation approaches utilizing optimization or planning can incorporate the noise metric as an additional objective.
A third avenue for future work is to integrate strategic separation approaches to limit the total number of aircraft that can pass through an airspace at a given time. Incorporating methods such as demand-capacity balancing with our real-time approach would allow the increase of traffic throughput of the airspace while retaining the safety and noise benefits of the RL policy.

\section*{Acknowledgments}
This work was partially supported by the National Aeronautics and Space Administration (NASA) University Leadership Initiative (ULI) program under project ``Autonomous Aerial Cargo Operations at Scale'', via grant number 80NSSC21M071 to the University of Texas at Austin. The authors are grateful to NASA project technical monitors and project partners for their support. Any opinions, findings, conclusions, or recommendations expressed in this material are those of the authors and do not necessarily reflect the views of the project sponsor.

\bibliography{sample}

\end{document}